\documentclass{kluwer}    
\usepackage{epsf}
\newcommand{\be}{\begin{equation}}
\newcommand{\ee}{\end{equation}}
\newcommand{\bea}{\begin{eqnarray}}
\newcommand{\eea}{\end{eqnarray}}

\begin{document}                                                                                   
\begin{article}
\begin{opening}         
\title{Origin of the torsional oscillation pattern of solar rotation}
\author{H.C. Spruit} 
\runningtitle{Torsional oscillation}
\institute{Max-Planck-Institut f\"ur Astrophysik, 85741 Garching,  Germany}
\date{}

\begin{abstract}
A model is presented that explains the `torsional oscillation' pattern of deviations in the solar rotation rate as a geostrophic flow. The flow is driven by temperature variations near the surface due to the enhanced emission of radiation by the small scale magnetic field. The model explains the sign of the flow, its amplitude and the fact that the maxima occur near the boundaries of the main activity belts. The amplitude of the flow decreases with depth from its maximum at the surface but penetrates over much of the depth of the convection zone, in agreement with the data from helioseismology. It predicts that the flow is axisymmetric only on average, and in reality consists of a superposition of circulations around areas of enhanced magnetic activity. It must be accompanied by a meridional flow component, which declines more rapidly with depth.
\end{abstract}

\end{opening}           

\section{Introduction}

The so-called torsional oscillation (Howard and LaBonte 1980) was discovered as a small time- and latitude-dependent modulation of the rotational velocity of the Sun as measured from Doppler shifts. In averages of the azimuthal component of the solar surface velocity field (synoptic maps) it appears as a band of  increased velocity that drifts towards the equator during the sunspot cycle, together with the magnetic activity. Its amplitude is of the order $\sim 5$ m/s, and is most prominent on the equatorward side of the main activity belt. 

The obvious connection of the flow with the solar cycle, and its relatively small amplitude, suggest that it may be a secondary effect, somehow caused by the magnetic fields that are the main manifestation of the cycle\footnote{If this is the case, the term `oscillation' would be somewhat misleading, as it suggests a cyclic variation due to some restoring force intrinsic to the flow itself.}. One possibility would be that it is driven by uncompensated Lorentz forces; this hypothesis has been put forward immediately after the discovery of the oscillation (Sch\"ussler, 1981), and has been the basis of several subsequent interpretations. An exception is Ulrich's (2001), which proposes that a hydrodynamic mode of the solar envelope must play an essential role, since the flow  is already present at activity minimum, when no active regions are present. The energy in the oscillation is small, however, compared with the inferred magnetic energy of the cycle, hence the suggestion in Ulrich et al. (2002) that the oscillation is in step with the cycle because it triggers the eruption of active regions. 

Instead of being a direct consequence of Lorentz forces, the oscillation could be the secondary result of a thermal effect of the cycle's magnetic fields. The field wound up by differential rotation in the interior of the convection zone could cast a `thermal shadow' on the layers above it by interfering with the efficiency of convection (Spruit, 1977; Parker, 1987). Gilman (1992) proposes that Coriolis forces acting on the downflow created by such a shadow would set up a pattern of differential rotation. The advantage of this idea is that it produces a concentration of the flow at the boundaries of the active latitude belt, as observed. It is not clear if thermal shadows would be of sufficient amplitude for this to work, especially if the shadowing takes place near the base of the convection zone. In a turbulent diffusion model the temperature changes due to shadowing are quite small, because of  the very high effective thermal conductivity of the convective envelope (Spruit, 1977). They are likely to be even smaller (Spruit, 1997) when the extreme asymmetry between upward and downward flows due to the density stratification of the envelope is taken into account, since this causes the upflows to be much more identical in entropy than in mixing length models (for results of recent numerical simulations see Stein and Nordlund (1998), Asplund et al. (2000). 

The bulk of the flux and energy of the solar magnetic field is believed to be located near the base of the convection zone (D'Silva and Choudhuri, 1993; Caligari et al., 1994; Fan et al., 1994). These authors have shown that the phenomenology of magnetic fields observed at the surface can be quantitatively understood as due to the eruption of the loops from this magnetic layer at the base, as in Leighton's (1969) classical interpretation of the solar cycle. Models in which the torsional oscillation is a consequence of magnetic fields therefore typically also place the source of the oscillation in the deeper layers of the convection zone. 

With only surface observations available, this prediction could not be tested but with the detailed measurements of the Sun's internal rotation made possible by helioseismology this has now changed. The torsional oscillation has been clearly detected in the variation of the rotation profile during the cycle (Woodard and Libbrecht, 1993b; Basu and Antia, 1998, Schou et al., 1998). Perhaps surprisingly, the oscillation appears to have its largest amplitude at the solar surface. In the most detailed results so far (Howe et al., 2000, Vorontsov et al., 2002) the pattern can be followed to a depth of about 100\,000 km. The systematic decline of the amplitude with depth, however, is hard to reconcile with an origin in the deeper layers of the convection zone.

The new interpretation proposed in this paper takes the apparent surface origin seriously. It proposes that the oscillation actually is i) a secondary consequence of the cycle's magnetic fields, and ii) is caused by the {\it thermal effect} which small scale surface magnetic fields of the cycle have on their surroundings, not the Lorentz force. As shall be demonstrated below, this provides a simple and easily quantifiable explanation for the oscillation. It agrees with most of its observed properties and makes detailed predictions. 

The key element of the theory is the observation that horizontal flows on length scales like those of the oscillation are dominated by the Coriolis force, hence approximately {\it geostrophic}, like large scale flows in the Earth's atmosphere and oceans.

\subsection{Relation of the oscillation to the magnetic field}

\begin{figure}
\mbox{}\hfill\epsfxsize 0.5\hsize\epsfbox{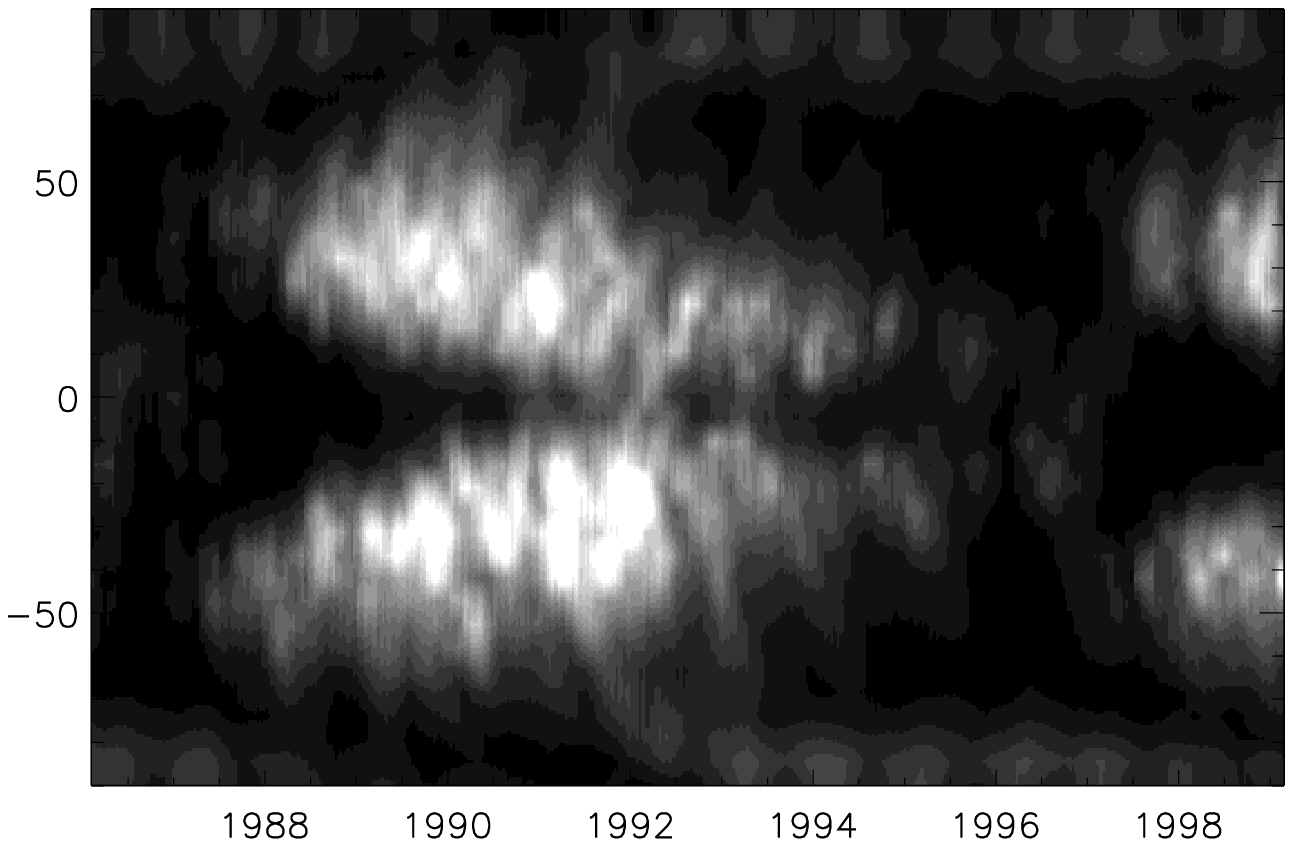} \hfill\epsfxsize 0.5\hsize \epsfbox{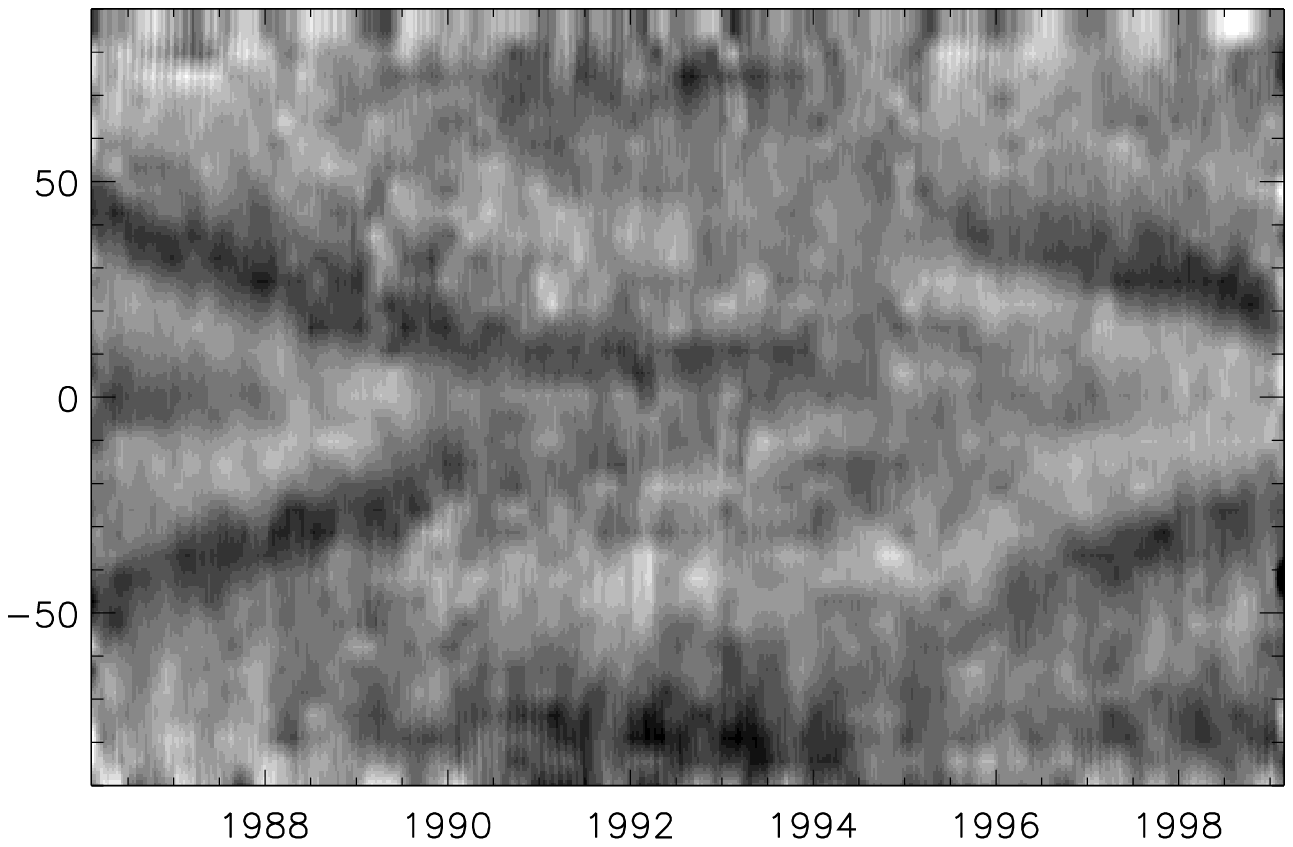} \hfill\mbox{}\break
\mbox{}\hfill\epsfxsize 0.8\hsize\epsfbox{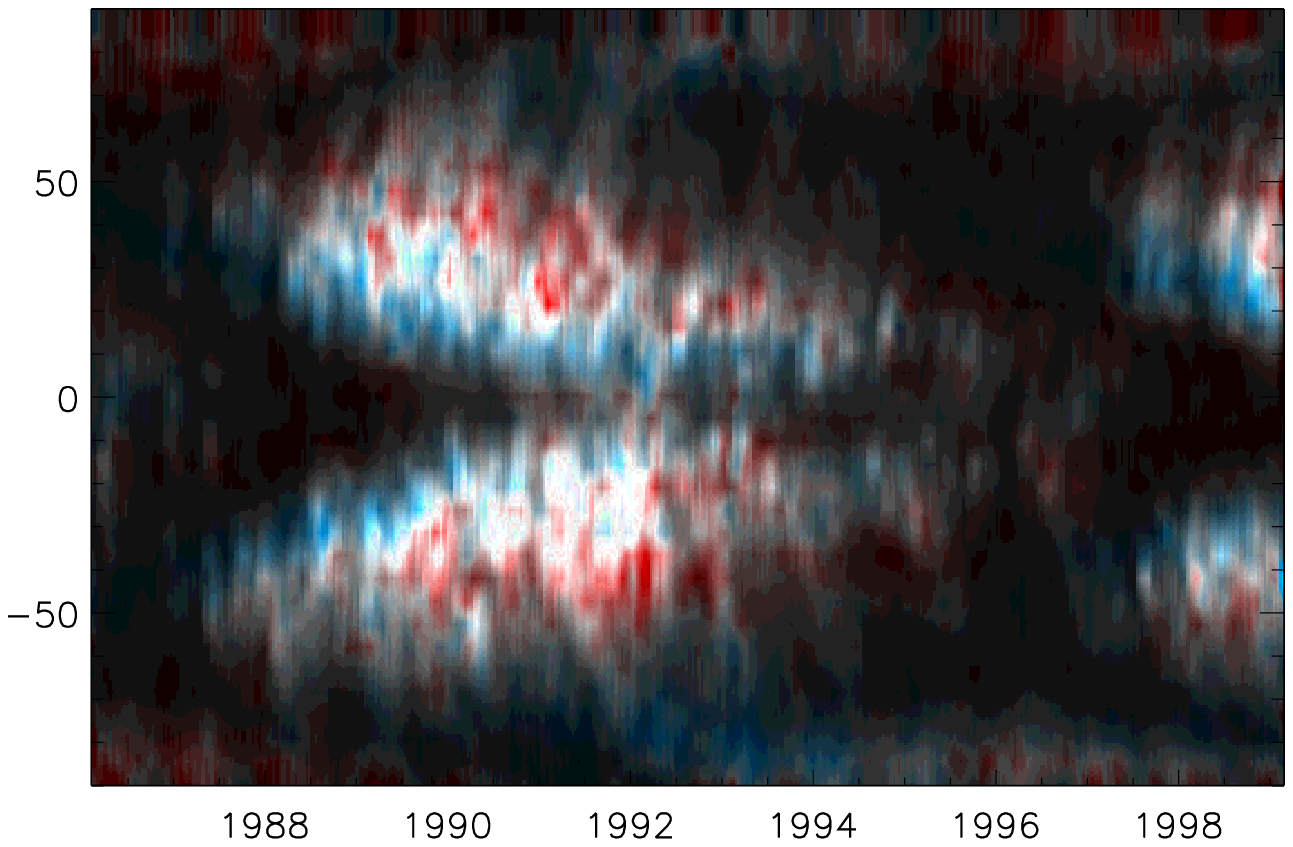}\hfill\mbox{}
\caption{Top panels: synoptic maps of the absolute value of the magnetic field (left) and azimuthal velocity residual (right) measured at Mt. Wilson Observatory (Ulrich, 2001). Bottom panel: composite map in which the absolute value of the magnetic field is shown in intensity and deviations in rotation velocity in color, with blue (red) for faster (slower) rotation than the mean. Color saturation codes for the velocity amplitude. Data have been averaged over 3 rotations.\label{relat}}
\end{figure}

The oscillation pattern travels with the main activity belt in a butterfly diagram, but its maximum does not coincide with that of the magnetic activity. To illustrate the well known relation between the field and the flow, I reproduce  in Figure \ref{relat} a synoptic map based on the observations made at Mt. Wilson from 1986 till 1999 (Ulrich 2001).

Deviations in rotation velocity are clearly associated with the magnetic activity, but avoid the center of the main belt of activity. As the lower panel shows, the velocity deviations correlate with the latitudinal {\it gradient of the magnetic activity}. On the leading (equatorial) side of the belts the gradient is strongest, and the flows most pronounced. On the trailing side, the flows are of opposite sign. The boundary of the activity belt is more diffuse on this side, and the flows are of lower amplitude. 

The representation in Fig. 1 is chosen to emphasize the relation which I want to draw attention to. It  somewhat deemphasizes a known problem with the oscillation, namely the fact that it starts before the magnetic field of the new cycle becomes noticeable on synoptic magnetograms. This problem is intrinsic to explanations that assume the cycle's magnetic field to be the cause of the zonal flows: the flows are quite strong already before the new cycle, as measured by the main magnetic activity indicators, sets in. I return to this problem in the discussion section.

The correspondence between velocity and gradient of the (absolute value of the) field strength suggests that the flows are geostrophic, like the large scale flows in the earth atmosphere and oceans. In these flows, there is an approximate balance between the horizontal pressure gradient and the Coriolis force, while the contribution of small scale (`turbulent') momentum transport processes is small on these length scales. The flow speed is therefore proportional to the horizontal pressure gradient. The direction of the flow is perpendicular to the horizontal pressure gradient, leading to the characteristic cyclonic and anticyclonic flows circling around low and high pressure systems, respectively (for a systematic treatment see Pedlosky, 1982). 

On the Sun, the pressure gradients driving geostrophic flows could be thermal, as in the geophysical case, or could have a magnetic origin. The magnetic field could cause flows directly, through the Lorentz force. A second possibility is an indirect, thermal effect they have on their surroundings; an example is the thermal shadow effect mentioned. Another is the well known surface effect of magnetic fields: by inhibiting convection (in spots) or facilitating radiative loss (in the small scale field), magnetic fields reduce or enhance the radiative cooling of the solar surface. This effect is entirely passive: the energy content of the magnetic structures is not affected, they only act as `valves' changing the radiative energy flux.

These thermal surface effects, though indirect, are not small. The excess emission due to small scale magnetic fields persists for as long as the magnetic structures are present on the surface. The total amount of excess energy emitted at the surface by a small flux tube during it life can therefore be much larger than its total magnetic energy content (e.g. Spruit et al. 1991). I show in this paper that this thermal effect produces flows of the observed nature and magnitude.

\section{Geostrophic flows caused by active regions}
\label{est}
The behavior of slow horizontal flows in a rotating fluid is governed by the Rossby number and, if the effects of viscous friction are important, the Ekman number. If viscosity can be ignored, the rate of change of the flow (left hand side of the Euler equation) can be compared with the strength of the Coriolis force. If the characteristic time scale $\tau$ of the flow is long, the Rossby number $Ro=1/(\Omega\tau)$ is small, and there is an approximate balance between the pressure gradient and the Coriolis force. This is called the geostrophic balance; it holds, for example, for the large scale flows associated with weather systems in the Earth's atmosphere. The latitudinal component of this balance is
\be 
\partial_y p=  - 2\rho v_x\Omega\sin\Lambda,\label{geob}
\ee
where $\Lambda$ is the latitude and $x,y$ are local cartesian coordinates with $x$ counted positive in the azimuthal direction (in the direction of rotation), and $y$ positive towards the North pole.

In flows on smaller length scales $L$, friction becomes important as well; its effects are measured by the Ekman number, $E=\nu/(\Omega L^2)$. Friction has to be taken into account when $E\gsim 1$. Small length scales are present at the boundaries of a large scale flow. The effects of viscosity in a rotating flow are thus typically observed in the form of {\em Ekman layers} at the boundaries of large scale flows.

To obtain typical values for the Rossby and Ekman numbers characterizing the torsional oscillation, take as the time scale $\tau$ the time for the pattern to change noticeably due to its drift in latitude. This would be a time scale of the order 1 yr, so the Rossby number is of the order $Ro\approx 0.01$. Assuming a kinematic viscosity $\nu\approx 10^{13}$ as a measure for the turbulent exchange due to granulation, we can determine the length scale $L_{\rm E}$ on which friction becomes important by setting $E\approx 1$. This yields $L_{\rm E}\approx 10$ Mm. On scales larger than this, the geostrophic balance holds; on scales of this order and smaller, the effects of friction have to be included.

\subsection{The active region belt: a low-pressure system}
\label{lowp}
The sign of the torsional oscillation, with an increased rotation rate on the leading and a reduced rate on the trailing side of the activity belts, is in the same sense as in a low-pressure system in the Earth's atmosphere. An interpretation of the oscillation as a geostrophic flow thus requires that the activity belts are zones of slightly lower pressure. 

By vertical hydrostatic balance, a  temperature perturbation extending from a depth $z=0$ to $z=d$ (counted positive into the Sun) causes a pressure perturbation $\delta p$ at $z=0$ given by
\be \delta p/p=\int_0^d\delta T/T {\rm d}z/H \approx n \delta T/T,\label{tem}\ee
where $H$ is the pressure scale height, and $n$ the number of pressure scale heights over which the perturbation extends. 

The observed sign of the flow thus requires that the subsurface temperature is lower in an active region. This may be the reason why the possibility of a geostrophic flow has not been considered much in explanations of the oscillation: the observed higher brightness of active regions would seem to argue for the opposite. 

In fact, the brightness of active regions produces just the right sign of the temperature perturbation. To see this if one has to suspend for a moment the intuitive notion that the brightness of active regions must be due to an enhanced temperature imposed from below. The effect of small scale magnetic fields on their immediate environment is a local increase of the heat flux escaping through the surface (Spruit 1977). A patch of solar surface with small scale fields has a slightly greater emissivity, a lower limb darkening, and it radiates a bit more like a black body than the normal solar surface. At any given horizontal surface the temperature in the magnetic structure is {\em lower} than average. That it nevertheless emits more is because the opacity in the structure is reduced, by the effect of the magnetic pressure. One result of this is that radiation is emitted from geometrically deeper layers than average, where the temperature is higher. Another is that when seen near the limb, the walls of the depression are seen. Even if those are not hotter than typical photospheric temperatures, they are seen face  on hence bright, as opposed to the surroundings, which are seen limb-darkened  (`bright wall effect').

The greater emissivity implies an {\it enhanced rate of cooling} of the gas exposed at the surface. As it cools at the surface the gas is swept into the intergranular lanes, so the cooling effect propagates down below the surface together with the intergranular downdrafts. The net result  of the presence of small scale fields at the surface is thus a slight temperature reduction below the surface. 

Apart from this theoretical prediction there are observational indications pointing in the same direction, most clearly from the p-mode frequency shifts during the solar cycle. A slight increase of the frequencies is observed with increasing activity. The $l$-dependence of the effect shows that it is caused mostly by a change in the wave propagation near the surface (Woodard and Libbrecht 1993a). Goldreich et al. (1991) noted that this is opposite to expectation if active regions are heated from below.  Higher temperatures would increase the propagation speeds, which acts in the right direction, but the thermal expansion of the envelope has to be included as well. This decreases the frequencies because of the increased path length, and dominates the frequency shifts because the expansion is linear in $T$, while the sound speed only increases as $\sqrt T$. 

In addition, a temperature increase, if it originates at a substantial depth, would give a different dependence of the frequency shifts on the degree $l$ of the modes. These conclusions have been substantiated with the much more accurate measurements of p-mode shifts with MDI (Dziembowski, Goode and Schou 2001; Dziembowski and Goode, 2002). The helioseismic observations thus rule out heating from below as the cause of the enhanced brightness in active regions.

The filling factor of the small scale magnetic fields is small, and so are the corresponding temperature changes. To see if the effect is sufficient by order of magnitude, consider a simple purely geostrophic estimate, based on eq. (\ref{geob}). If $L$ is the length scale of the perturbation in the latitudinal direction, the relative pressure perturbation is related to the flow speed $v$ by
\be \delta P/P=2\sin\Lambda {v\Omega L\over c_{\rm s}^2}, \label{relp}\ee
where $c_{\rm s}=(p/\rho)^{1/2}$ is the (isothermal) sound speed. With $v\sim 5$ m/s,  $L$ of the order of the width of the belt of activity, $L\sim 3\,10^{10}$ cm, and $c_{\rm s}\sim 10^6$ cm/s:
\be \delta P/P\sim 5\,10^{-5}. \ee
This small pressure effect can be produced by a temperature change of the same order or less (cf eq.\ \ref{tem}). 

This estimate of the flow speed does not take into account the effect of viscosity, however. The pressure difference needed for a given velocity will turn out the be significantly higher if friction is taken into account. This is done in the more detailed model developed below. For the moment, the estimate suffices to show that the temperature and pressure effects associated with geostrophic flows on the Sun can be quite small.

\section{Quantitative model}

To quantify these ideas better, a model is needed for the depth dependence of the temperature perturbations, as well as a model for the stratification of the convection zone.

\subsection{Stratification}
\label{strat}
The density in the convection zone varies over several orders of magnitude over the range of depths of interest here (several 10 Mm), hence its stratification has to be taken into account for a meaningful quantitative estimate. If it were just the stratification of density that is of importance, a simple polytropic stratification would be appropriate. But since the flows involve differences in temperature, the stratification of entropy has to be treated a bit more realistically than just using a simple isentrope. I adopt here the `pseudopolytropic' model used before in Spruit (1982, 1991). As in a standard polytrope, the temperature in this approximation varies linearly with depth into the Sun:
\be T=T_0\zeta, \qquad \zeta=1+z/z_0,\ee
where $z=0$ at the photosphere, and $z_0$ a fitting constant to be determined.
A good fit to the stratification of the mean density in a mixing-length model of the convection zone is
\be \rho=\rho_0\zeta^2, \ee
if $z_0$ is chosen of the order 
\be z_0\approx 4.5\, 10^7.\ee
The gas pressure 
\be P=P_0\zeta^3, \ee
is in hydrostatic equilibrium with $P_0=z_0\rho_0g/3$. The logarithmic temperature gradient is
\be {{\rm d}\ln T\over {\rm d}\ln P}=\nabla=1/3.\ee
This stratification corresponds to a polytropic index $n=2$, instead of the value $1.5$ for the isentropic stratification of a fully ionized gas. The value $n=2$ gives a better overall fit to the solar convection zone. This reflects the fact that ionization is partial over a significant fraction of its depth, in particular the first 10--50 Mm that are important here. 

Also because of partial ionization, the adiabatic gradient $\nabla_{\rm a}$ is not a constant in the  convection zone. The `pseudo' part of the pseudopolytropic model now consists of manipulating $\nabla_{\rm a}$ a bit. I make it a function of depth, such that the superadiabatic gradient $\nabla-\nabla_{\rm a}$ has a depth dependence which approximately fits the mean stratification of entropy in the envelope,
\be G\equiv\nabla-\nabla_{\rm a}=G_0 \zeta^{-2}\label{del}, \qquad {\rm with}\quad G_0\approx 0.25.\ee
In this way, a realistic entropy gradient can be included without giving up the algebraic convenience of a polytropic model.

\subsection{The temperature perturbation below the surface}
From the observed emissivity changes at the surface we can estimate the enhanced rate of cooling in active regions. In order to calculate what this implies in terms of temperature changes {\em below} the surface, a model is needed for the downward propagation of the cooling effect in the intergranular lanes. This depends on the details of mixing and entrainment between downdrafts and upflows. These are known in sufficient detail from numerical simulations of the upper few Mm of the convection zone (Stein and Nordlund 1998, Asplund et al. 2000), but are increasingly uncertain with increasing depth. 

These same mixing processes, however, also determine the mean stratification of entropy below the surface. This mean stratification is well determined. In numerical simulations for example, one finds that, though the downdrafts generated by cooling at the surface vary greatly in their value of the temperature contrast $\delta T(z)$, the {\it average} stratification of entropy is in fact reasonably close to that of a mixing-length model (Rosenthal et al. 1999). This stratification is also well determined observationally from helioseismology. It now turns out that the mean stratification of entropy is all that is needed to compute the effect that a small change in surface cooling has on the mean temperatures below the surface. This removes the uncertainty associated with mixing between upflows and downdrafts. To show this I consider two models for the mixing process. In the first, the scenario suggested by the numerical simulations mentioned is used. The result is then shown to be almost equivalent when a mixing length or turbulent diffusion model is used.

A downdraft generated as an intergranular lane at the surface with an entropy $S_{\rm i}$ is compressed to a smaller volume by the rapidly increasing pressure. This is offset by mixing with the nearly isentropic upflows surrounding it, which increases the filling factor of the downdrafts  again. In the process, however, the entropy difference with the surroundings decreases. The result is a mean entropy $\bar S(z)$ which increases with depth in such a way that the difference between this mean and the entropy $S_0$ of the upflows, $\Delta S(z)=\bar S(z)-S_0$, decreases with depth (producing the so-called `superadiabatic' layer). 

Horizontal pressure equilibrium is a good approximation in the downdrafts, and the entropy difference $\Delta S$ at a given pressure is approximately proportional to $\Delta T/T$ (exactly so for an ideal gas with constant $\gamma$). This difference $\Delta T/T$ between the mean temperature and that of the adiabatic upflows is a known function of depth (for example from helioseismology). In our pseudopolytropic approximation, it is found from (\ref{del}) by integration:
\be \Delta T/T=\int G {\rm d}\ln P={3\over 2}G_0\zeta^{-2}.\label{Del}\ee

In order to find the depth dependence of a small {\em perturbation} caused by a small change in the mean temperature of the downdrafts, I now assume linearity in the perturbation. Thus I  neglect the effect of this small change on the dynamics of the downdrafts, i.e. treat them as passively carried with the downflows. The depth dependence of the temperature perturbation is then given by (\ref{Del}):
\be \delta T/T=\epsilon\zeta^{-2},\label{dt}\ee
where $\epsilon\equiv (\delta T/T)_0$ is the value just below the photosphere (optical depth of a few, such that the effects of radiative transfer are already small).

In a conventional mixing length model, the depth dependence of the superadiabatic gradient is governed by the mixing length-to scale height ratio and the heat flux. By linearizing the dependence of the heat flux on the superadiabatic gradient in this model, the effect of a small change in heat flux can be calculated. If horizontal turbulent transport is neglected, the superadiabatic gradient would just increase by the same factor at all depths in response to an increase in heat flux. One verifies that this would, to within a numerical factor of order unity, yield the same result as (\ref{dt}). The main differerence thus lies in a possible contribution of horizontal diffusion of heat. Since the width of the active latitude belt is of the same order as the depth of the convection zone, the effect of such horizontal diffusion will also add only a factor of order unity. 

The depth dependence of the temperature effect due to enhanced surface emissivity is thus rather independent of the specific model for convective heat transport used. With sufficiently detailed helioseismic observations of the flow, however, it might be possible to distinguish between the predictions made by different models.

\subsection{First estimate: Purely geostrophic approximation}
\label{geos}

With this determination of the temperature effect, the expected geo\-stroph\-ic flow amplitude can be computed as a function of depth. I do this first in the inviscid approximation; this is then improved in the next section. 

By vertical hydrostatic equilibrium\footnote{One may wonder if vertical hydrostatic equilibrium of this temperature perturbation is compatible with the {\it horizontal} pressure equilibrium that holds for the individual downdrafts. This is a matter of length scale (or Ekman number). On the large horizontal scales considered here, the horizontal balance is between pressure perturbation and Coriolis force: the geostrophic balance. On the small granulation scale, the geostrophic balance breaks down, and is replaced by pressure equilibrium.} we get the pressure perturbation associated with $\delta T$:
\be 
\delta\ln P=\int \delta T/T\, {\rm d}z/H={\epsilon z_0\over 2H_0} \zeta^{-2}, \label{delp}
\ee
where $H_0$ is the surface pressure scale height $H_0\approx 1.5\,10^7$cm.
From the geostrophic balance (\ref{relp}), and the estimate for $\epsilon$ given in  \ref{est}, this yields the azimuthal velocity amplitude of the geostrophic flow:
\be v=v_{\rm g}\zeta^{-1},\qquad v_{\rm g}={\epsilon z_0\over 2H_0}{c^2_{{\rm s}0}\over 2\Omega L\sin\Lambda},\label{vest}\ee
where $c_{{\rm s}0}\sim 10$ km/s is the isothermal sound speed near the surface.

For a quantitative estimate, a plausible value is needed for the amplitude $\epsilon$ of the temperature change below the surface. I do this first from a theoretical consideration of small flux elements, and then check this estimate against the bolometric changes in the solar flux known from the measured irradiance variations.

A small flux element (`tube') channeling additional radiation through the surface (Spruit 1976, 1977) can have a bolometric contrast of some 10\%. This is the total effect, including the area of reduced temperature (`dark ring') immediately surrounding it, and normalized by the cross section of the element. Very small elements can have significantly larger values, while in bigger ones (knots, pores) the contrast is actually negative, as in sunspots. The value of 10\% agrees with observations of the smallest elements where this contrast can be measured directly. An a priori estimate of the effect of the small scale field in the activity belt thus depends on the {\em distribution of sizes} of magnetic elements, which is poorly known (see however Spruit and Zwaan, 1981). This is at the moment the main uncertainly in comparing the theory to observations. 

Better known is the filling factor of the small scale magnetic field. Averaged over the active latitudes, this is of the order 1\%, corresponding to an average (absolute) surface field strength of 10 G. The 10\% effective contrast of the flux elements thus translates into a mean surface flux increase $\delta F/F\approx 10^{-3}$. This can be compared with the irradiance variations measured at Earth during the solar cycle (e.g. Fr\"ohlich and Lean, 1998). These show a peak-to peak variation of about $7\, 10^{-4}$. Assuming that the main contribution of this signal comes from the active latitudes, with a width $\Delta\Lambda\approx 30^\circ$, this corresponds to a variation in surface flux within the active latitudes of $\delta F/F\approx1.5\,10^{-3}$. This is quite comparable to the theoretical estimate above. I assume a value $\delta F/F= 10^{-3}$ in what follows.

The number for the difference in cooling rate needs to be translated into a temperature change below the surface. The radiation emitted by the granulation flow, from the time radiative loss first becomes important till the time the flow descends back in the intergranular lanes, produces the entropy difference between the downdrafts and the upflows. This is known from observations and from numerical simulations of granulation; it corresponds to a relative temperature change (at a given pressure) of order unity (from about 11\,000 K to 6000 K). A change $\delta F/F$ in the emitted radiation thus yields a temperature change in the downdrafts of the same amount, $\epsilon\approx \delta F/F$. Thus our estimate for the average temperature effect due to flux tubes in the activity belt is 

\be \epsilon\approx 10^{-3}.\ee

Taking for the $L$ the approximate width of the sunspot belt, $L\approx 3\, 10^{10}$, and this value for $\epsilon$ we get $v\approx 5$m/s at $z=10$ Mm. This is comparable with the values observed helioseismically at this depth. 

At the surface, however, the estimate predicts much larger values than observed, around $v_{\rm g}=100$ m/s. As discussed in \ref{lowp}, this is because we have neglected friction, which is important over the first $20$ Mm or so. The surface is dominated by an {\em Ekman layer}. This is the subject of the next section.

\section{Ekman layer}

In this section we derive the flow which develops under the influence of a steady horizontal pressure perturbation near the surface. The effect of mixing by the small scale granulation flows is modeled by an isotropic kinematic viscosity coefficient. The resulting model is very similar to the standard model for an Ekman flow in geophysics. Due to the viscosity, an Ekman layer develops near the surface, which modifies the flow compared with a purely geostrophic approximation.

The main difference with respect to the geophysical case is the very strong density stratification in the solar envelope. I take this into account using the analytic stratification model presented in section \ref{strat}. 

\subsection{Equations}

The equation of motion for steady flow of a viscous fluid in a rotating system is
\be 0= -\nabla p +2\rho{\bf v}\times{\bf \Omega}+ {\bf \tau},\ee
where ${\bf \tau={\rm div}\sigma}$ is the viscous force, in terms of the viscous stress tensor $\sigma_{ij}$. The value of the kinematic viscosity coefficient, from a mixing length model for the convection zone, is 
\be \nu\approx{1\over 3}l v\approx 10^{13}{\rm ~~cm^2/s},\ee
relatively independent of depth. The dynamic viscosity $\eta=\rho\nu$ is thus a very strongly increasing function of depth. The viscous stress in such a stratified fluid is (e.g. Landau and Lifshitz, 1982)
\be 
\sigma_{ij}=\rho\nu ({\partial v_i\over \partial x_j}+{\partial v_j\over \partial x_i}) +\rho(\chi-{2\over 3}\nu)\delta_{ij}{\rm div}{\bf v},
\ee
where $\chi$ is the (kinematic) compressional viscosity coefficient. Both $\nu$ and $\chi$ are uncertain, depending on the details of the turbulence model used, but of similar order of magnitude. Since the essence of the flow studied depends mostly on the shear viscosity, I simplify the equations by assuming $\chi={2\over 3}\nu$ so that the compressional viscous terms disappear.

The effect of the Coriolis forces depends on latitude. For the present estimates, it is sufficient to consider the situation near the North pole, so that the rotation vector is vertical and pointing out of the Sun. Use a local cartesian coordinate system with azimuthal direction $x$, latitudinal direction $y$ and vertical $r$ (counted positive out of the Sun). This vertical coordinate has the opposite sign from that used for the stratification model of \ref{strat}, i.e. $r=-z$. This is done here so that the rotation rate $\Omega$ will be a positive number. 

The horizontal equations of motion for an axisymmetric flow (independent of $x$) are then
\bea 
-\partial_y p\, P -2\rho v_x\Omega  +2\rho\nu\partial_{yy}v_y+\partial_r[\rho\nu(\partial_r v_y+\partial_y v_r)]=&0, \cr
 2\rho v_y\Omega+\partial_r(\rho\nu\partial_rv_x)=&0,\label{eqmo}
\eea
where $p$ is now the relative pressure perturbation $\delta P/P$, and $P(r)$ the mean pressure stratification. 

To represent the dependence of the driving pressure gradient on latitude it is sufficient to take a simple sinusoidal form with wavenumber $k$. The resulting flows then also have a harmonic dependence on latitude with this wavenumber, and we can replace all latitudinal gradients by
\be \partial_y=i k.\ee
Further simplification is possible by noting that the depth of the Ekman layer, of the order $L_r=10$ Mm (see \ref{est} above), is small compared with the latitudinal scale of the flows, of the order $L_{\rm h}\sim 100$ Mm or more.  Expanding eqs.~\ref{eqmo} systematically in the small parameter $L_r/L_y$, we get
\bea
-ikp P-&2\rho v_x\Omega+\partial_r(\rho\nu\partial_r v_y)=0,\cr
       & 2\rho v_y\Omega+\partial_r(\rho\nu\partial_r v_x)=0.
\eea
Due to this expansion in  $L_r/L_y$, the vertical velocity has disappeared from the equations. Hence neither the vertical equation of motion nor the continuity equation are needed to solve the problem (if needed, the vertical velocity can be reconstructed afterwards from the solution by using the continuity equation).

Four boundary conditions are needed. Since the driving force vanishes with depth (eq.\ \ref{delp} above), the velocity also vanishes at large depths: 
\be v_y, v_x\rightarrow 0 \quad (z\rightarrow \infty).\ee
In the atmosphere, the granulation velocity and hence the viscosity vanishes with increasing height. This implies that the appropriate boundary conditions at the surface $z=0$ are stress-free:
\be \partial_r v_y=\partial_r v_x=0 \quad (z=0).\ee
To see this, approximate the decrease of the viscosity with height as a jump at $z=0$. Across this jump, the $r-y$ stress must be continuous (because there are no surface forces, only volume forces). Since the viscosity vanishes above the jump, this stress vanishes. Just below the surface, this implies $\partial_r v_y=0$, because the viscosity is finite there. The boundary condition to be used is thus stress-free. 

Using the analytic stratification model from the previous section, and the depth dependence of the relative pressure perturbation from the model of section \ref{strat}, the equations of motion are, in terms of the dimensionless depth coordinate $\zeta=1+z/z_0=1-r/z_0$:
\bea 
-i v_{\rm g}\zeta -&\zeta^2v_x+E\partial_\zeta(\zeta^2\partial_\zeta v_y)=0,\cr
                 & \zeta^2v_y+E\partial_\zeta(\zeta^2\partial_\zeta v_x)=0.\label{eqz}
\eea
In these equations,
\be E={\nu\over 2\Omega z_0^2}\ee
is now a `surface value' of the Ekman number, i.e. the value based on the length scale $z_0$ of the stratification near the surface. From the model stratification of section \ref{strat} and with a kinematic viscosity $\nu\approx 10^{13}$ cm~s$^{-2}$ as before,  we have $E\approx 800$. The constant
\be v_{\rm g}={\epsilon z_0\over 2H_0}{kP_0\over 2\rho_0\Omega}\ee
(with the dimension of a velocity) is the purely geostrophic velocity amplitude introduced in section \ref{geos}. It measures the strength of the driving force, in terms of the amplitude $\epsilon$ of the surface cooling effect.

\subsection{Solution}
\label{sol}
By inspection, one finds that the solutions of the homogenous part ($_{\rm H}$) of eqs (\ref{eqz}) which satisfy the BC at $\zeta\rightarrow\infty$ are of the form
\be (v_{x,y})_{\rm H}\sim \zeta^{-1}\exp[{\zeta\over (2E)^{1/2}}(-1\pm i)], \ee
while a particular solution ($_{\rm I}$) is
\be  v_{y{\rm I}}=0, \qquad v_{x{\rm I}}=-iv_{\rm g}\zeta^{-1}.\ee

Replacing the latitudinal dependence $e^{iky}$ by its real equivalent, so that 
\be p\sim \cos(ky),\ee
and applying the stress-free boundary conditions at the surface $\zeta=1$, the solution is found to be 
\be v_x=v_{\rm g}\sin(ky)\zeta^{-1}[1-\cos(\zeta/\lambda)e^{-\zeta/\lambda}],\ee
\be v_y=v_{\rm g}\sin(ky)\zeta^{-1}\sin(\zeta/\lambda)e^{-\zeta/\lambda},\ee
where
\be \lambda=(2E)^{1/2}\approx 40.\ee
The direction of the flow rotates with depth, while its amplitude decreases. It is similar to the classical `Ekman spiral' as is produced, for example,  by wind stress on the surface layers of the oceans (for a detailed discussion see Pedlosky 1982, Ch. 4.10). One difference with the geophysical case is the additional factor $\zeta^{-1}$, which is a consequence of the density stratification. In addition, since the driving force is present throughout the volume rather than only at the surface, the flow declines more slowly with depth than in a classical Ekman layer.

Since $1/\lambda$ is a rather small number, the velocity at the surface ($\zeta=1$) is approximately
\be v_{y0}\approx v_{x0}\approx {v_{\rm g}\over\lambda}\sin(ky). \ee
The velocity at the surface thus makes an angle of  nearly 45$^\circ$ with respect to the direction of the driving force, just as in the classical Ekman spiral. 

With the value of the surface cooling parameter $\epsilon$ as estimated in section \ref{geos}, we have $v_{\rm g}\approx 250 $m/s, and $v_{y0}=v_{x0}\approx 6$m/s. The solution with these parameter values is shown in Fig. 2.  The oscillatory behavior typical of an Ekman layer is still present, but much less noticeable than in the geophysical case. This is due to the additional factor $\zeta^{-1}$ in the depth dependence.

\begin{figure}
\mbox{}\hfill\epsfxsize 0.8\hsize\epsfbox{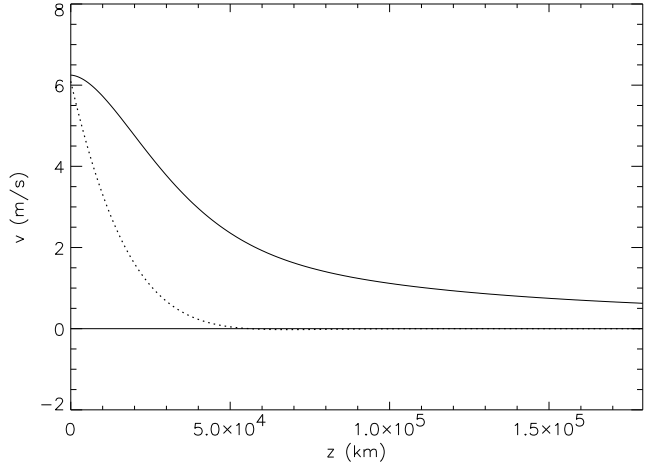}\hfill\mbox{}
\caption{Depth dependence of the Ekman flow. Solid: azimuthal (zonal) component, broken: latitudinal component. }
\end{figure}

\section{Interpretation and predictions}

The amplitude $v_{\rm g}$ of the flow at the surface is now smaller than the purely geostrophic estimate (\ref{vest}) by the factor $1/\lambda\approx 0.025$. For the same value of the driving parameter $\epsilon$ it is now $v_{\rm g}\approx 6$ m/s, in excellent agreement with the surface Doppler measurements (e.g. Ulrich 2001). The model predicts a velocity amplitude that declines smoothly with depth, to about 1/3 of  its surface amplitude at a depth $z\approx 60$ Mm. At the base of the convection zone, the amplitude is still about 10\% of the surface value. The flow thus extends over much of the convection zone, though with reduced amplitude compared to the surface, in agreement with the helioseismic observations (Howe et al., 2000; Vorontsov et al. 2002). 

The velocity pattern produced by the model is in agreement with the observation that the flow is concentrated at the {\em boundaries} of the main activity belt (see Fig. 1). The {\em sign} of the flow agrees, provided we accept that the small flux elements in an active region acts as `leaks' by which an enhanced heat flux passes through the surface, as discussed in section \ref{lowp}. The consequence of this leak is additional {\em cooling} below the surface in active regions, the opposite of what would be the case if they were heated from below. 

The model makes a few testable predictions. 

\subsection{predictions}

\subsubsection{Meridional component of the Ekman flow}
A generic feature of an Ekman spiral is the direction of the flow at the surface, which is at an angle of nearly 45$^\circ$ to the direction of the driving force. In our application, this means that there must be a flow directed from the edges towards the center of the main activity belt. Its amplitude must the same as that of the torsional oscillation. This feature has, in fact, been observed (Komm, Howard and Harvey 1993, Komm 1994). From data obtained at Kitt Peak these authors find a pattern of meridional flow in just this sense, directed towards the center of the active latitude zone at an amplitude of about 5 m/s. 

\subsubsection{Depth dependence}
A further prediction is that the direction of the flow changes with depth. As Fig. 2 shows, the latitudinal velocity component is comparable to the zonal component only in a thin layer near the surface. It declines much more rapidly than the zonal component and disappears below about 30 Mm. This can probably be tested with helioseismic data. Some results relevant in this context have already been given by Gonz\'alez Hern\'andez \& Patr\'on (2000).

\subsubsection{Nonaxisymmetric structure}
The model also predicts that the flows induced by active region cooling are axisymmetric only on average. Activity patches with scales sufficiently large compared with the Ekman layer depth $z_{\rm E}=z_0\lambda\approx 20$ Mm, and sufficiently long lived compared with the rotation period, should each have a cyclonic circulation around it (i.e. directed with respect to the rotation axis in the same sense as a low-pressure system on Earth). These flows should peak in the outer parts of the patch, and be accompanied by a flow towards its center with approximately the same surface amplitude. With depth, this flow would become more nearly geostrophic . On smaller scales, between $10$ and $100$ Mm, the flow speeds produced by the same field strength would be higher because of the stronger horizontal gradients, though the smaller scales would also make such features harder to detect. 

Some nonaxisymmetric structure in the flow has, in fact, already been detected. Ulrich (1998, 2001) shows the existence of long-lived azimuthal orders $m\lsim 8$. 

\subsubsection{Speed of downward propagation}
The model predicts that the flows start at the surface, together with their cause: the surface cooling due to the small scale magnetic fields. They would propagate down together with the downdrafts, so that it would take some time for the flows to develop at greater depths. The predicted time delay is small, however. The speed of downdrafts is of the order 1 km/s at the surface, decreasing with depth due to mixing and entrainment. Since the filling factor of the downflows in strongly stratified convection is less than the $50$\% assumed in mixing length models, the downdraft speed is larger at all depths than the mixing length value. Assuming as a conservative value for the mixing length velocity in the lower half of the convection zone of about 100m/s, the time it takes for the effects to propagate to a depth of 100 Mm is less than $10^6$s, or less than about 2 weeks. 

Compared with the activity cycle, the flows are thus set up essentially instantaneously, and the prediction is that they closely track the distribution of small scale fields at the surface. This can be tested in principle, but a precise measurement is complicated by the fact that the mean rotation in the convection zone is known only with a finite accuracy. Since the mean rotation rate varies mostly with latitude, a small change in the assumed depth dependence will introduce latitudinal kinks into originally vertical structures in the pattern of rotation. The results of Vorontsov et al. (2002, their Figure 3) in fact show such sideways displacements, of both signs (directions of apparent propagation both upward and downward). The effect appears to be small compared with the latitudinal drift of the oscillation with the cycle, however.

\section{Discussion}

I  have presented a geostrophic flow model for the torsional oscillation, which identifies it as a consequence of enhanced surface cooling by the small scale magnetic field in active regions. Its achievements are:
\begin{itemize}
\item{} The model is quantitative and makes detailed predictions. 
\item{} It explains the sign of the oscillation and its concentration along the boundaries of the active latitude belt.
\item{} Using observed values for the enhanced emissivity of active regions, and a standard value for the turbulent viscosity due to granulation, it reproduces the correct amplitude of the flow.
\item{} It explains why the flow has its maximum amplitude at the surface, and reproduces the approximate depth dependence as measured with helioseismology.
\end{itemize}
The accuracy of the model can still be improved with better input data. The most critical data it needs are the bolometric brightness contrast of active regions, more generally: all small scale magnetic fields including ephemeral regions and the network. The surface observations required for this currently have limited accuracy, or they rely on proxy indicators like Calcium emission. This will probably improve in the future, since these measurements are the same as are needed for an improved identification of the cause of  solar irradiance variations.

The main difficulty the model faces is that the correct correlation of the oscillation with magnetic activity (as seen in magnetograms) breaks down around minimum activity, when the flow is already strong but little evidence for magnetic fields of the new cycle is seen. This problem is addressed below in \ref{early}.

\subsection{Active regions as cooling agents}

The model  is based on the fact that active regions must be areas of lower temperature below the surface. This is perhaps somewhat counter to the prevailing intuition. One might feel that, since active regions are brighter, they should have higher temperatures. This intuition is also somewhat rooted in the idea that the energy flux of the convection zone comes from below, so that it is natural to associate higher surface temperatures with higher heating from below. 

This intuition is misleading, however. In a steady state, the heating from below is equal to the cooling at the surface by radiation into space, and it is hard to say which of the two is more fundamental in determining the structure of a convective envelope. In a time dependent situation, the two are generally not equal. Which of the two agents, heating from below and cooling at the top, provides the most appropriate physical view when considering {\em changes} in the convection zone depends on the {\em time scale} and location of the phenomenon studied. On the time scale of stellar evolution, it is the change in conditions at the center that determine the changing luminosity of the star. The thermal time scale of the star, $\approx 10^7$ yr, is very long compared to the phenomena of solar activity, however. Even the thermal time scale of the convection zone alone, $\sim 10^5$ yr, is very long in this respect. 

Because of this, changes in surface brightness due to variations in conditions in the deeper layers of the convection zone, though not impossible, are much harder to bring about than changes in brightness caused by surface effects. The darkness of sunspots due to the suppression of convection by their magnetic field is a well-known example of such a surface effect. The opposite of this effect happens in small scale magnetic fields, which act as leaks in the surface through which an increased heat flux escapes from the convection zone (Spruit 1977). These effects can be viewed simply as modulations of the effective emissivity of the solar surface, i.e. changes in the surface heat flux at {\it fixed} thermal conditions in the {\it bulk} of the convection zone. 

The large differences in time scale and energy required for thermal changes in the bulk compared with changes at the surface is due to the enormous stratification of the density through the convection zone, by a factor of $10^6$. This has no equivalent in the laboratory- and kitchen analogs of convecting fluids on which conventional intuition is based.

\subsection{Flows: a proxy for surface temperature variations}
As shown above (\ref{sol}) the velocity amplitude produced by a 0.1\% change in surface  emissivity is of the order 5 m/s, for a nominal horizontal length scale of $3\,10^{10}$ cm. Velocity amplitudes of this magnitude are easily detected with current Doppler and helioseismic measurements. The brightness changes themselves are much more difficult to observe directly. So far, the sensitivity of  the order 0.03\% needed for such a detection has been achieved only with disk-integrated measurements like the ACRIM irradiance data. Surface flows like the torsional oscillation may therefore be a more sensitive way of detecting surface temperature changes associated with the solar cycle.

\subsection{The early start of the flow pattern}
\label{early}
A difficulty with the model presented here is the strength of the observed oscillation at high latitudes, before the beginning of the spot cycle. In synoptic magnetograms, there is very little evidence of magnetic fields at these latitudes and phase in the cycle (e.g. Ulrich, 2001, Ulrich et al. 2002). The model presented here predicts that substantial magnetic flux is in fact present there, but in a form which somehow escapes detection in synoptic magnetograms. 

This is not a new suggestion. Significant coronal activity at high latitudes before the beginning of the spot cycle has been noticed already in the 60's, in particular in the Pic du Midi observations of Trellis et al. (cf.\ Leroy and Trellis, 1974). It has also been consistently seen in the coronographic observations at Sacramento Peak (Altrock, 1986; Guhathakurta et al., 1993). Their possible connection with the torsional oscillation was first proposed by Wilson et al. (1987).

The p-mode frequency shifts during the activity cycle show a similar effect: the latitude dependence of the sources derived from these shifts agrees well with the surface distribution of magnetic fields (Antia et al. 2001), except at high latitude, where the effect is stronger than expected from magnetograms (Moreno-Insertis and Solanki, 2000). The inferred high-latitude magnetic fields peak around activity minimum (Dziembowski and Goode, 2002; Goode et al., 2002).

The form in which these high latitude fields can exist is significantly constrained by their absence in synoptic magnetograms. They would have to exist in the form of small scale flux with polarities nearly balancing within the pixel scale of the synoptic maps. High-resolution observations at the relevant latitudes around solar minimum could test this possibility. Ephemeral active regions, known to have a wider latitude distribution than regular active regions and to be most prominent at activity minimum (Harvey et al. 1986, Harvey 1989), may be a manifestation of these fields (Wilson et al. 1987). It is also possible that magnetograms are less sensitive to magnetic flux in very small scale form, due to radiative transfer effects (e.g.\ S{\' a}nchez Almeida 2000). 

\acknowledgements
I thank Dr.\ Roger Ulrich for extensive discussions and for kindly making the Mt.\ Wilson observations of magnetic fields and torsional oscillations available in numerical form (used here for Fig. 1). Suggestions and critical comments by Dr. Martin Woodard have been valuable for improving the presentation.


\end{article}
\end{document}